\documentclass[twocolumn,preprintnumbers,amsmath,amssymb,superscriptaddress,unsortedaddress,byrevtex]{revtex4}
\usepackage{graphicx}% Include figure files
\usepackage{dcolumn}% Align table columns on decimal point
\usepackage{bm}% bold math
\usepackage{subfigure}

\begin{document}

\preprint{APS/123-QED}

\title{A Zero-Gravity Instrument to Study Low Velocity Collisions of
  Fragile Particles at Low Temperatures}

\author{D. M. Salter}%
\surname{Salter}
%\email[Author to whom correspondance should be addressed; electronic mail: ]{parabolic@strw.leidenuniv.nl}
%\affiliation{Leiden Observatory, Leiden University, Netherlands}
\affiliation{Leiden Observatory, Leiden University, Postbus 9513, 2300
  RA, Leiden, Netherlands}

\author{D. Hei$\ss$elmann}%
\affiliation{Institute for Geophysics and Extraterrestrial Physics,
  University of Braunschweig, Institute of Technology,
  Mendelssohnstra$\ss$e 3, 38106, Braunschweig, Germany}

\author{G. Chaparro}
\affiliation{Leiden Observatory, Leiden University, Postbus 9513, 2300
  RA, Leiden, Netherlands}
\affiliation{Kapteyn Astronomical Institute, University of Groningen,
  Postbus 800, 9700 AV, Groningen, Netherlands}
%\affiliation{Kapteyn Astronomical Institute, University of Groningen, Netherlands}
% \altaffiliation[Also affiliated with ]{Leiden University,
% Netherlands}
%\affiliation{Leiden Observatory, Leiden University, Netherlands}

\author{G. van der Wolk}
\affiliation{Kapteyn Astronomical Institute, University of Groningen,
  Postbus 800, 9700 AV, Groningen, Netherlands}

\author{P. Rei$\ss$aus}
\affiliation{Space Technology \& Applications Division, Kayser-Threde,
  GmbH, Wolfratshauser Stra$\ss$e 48, 81379, Munich, Germany}

\author{A. G. Borst}
\affiliation{Dutch Space B.V., Postbus 32070, 2303 DB, Leiden, Netherlands}

\author{R. W. Dawson}
\affiliation{Department of Physics SUPA (Scottish Universities Physics
  Alliance), University of Strathclyde, 107 Rottenrow East, Glasgow G4
  0NG UK}

\author{E. de Kuyper}
\affiliation{Fijn Mechanische Dienst, Leiden Institute of Physics,
  Leiden University, Postbus 9513, 2300 RA, Leiden, Netherlands}

\author{G. Drinkwater}
\affiliation{Department of Physics SUPA (Scottish Universities Physics
  Alliance), University of Strathclyde, 107 Rottenrow East, Glasgow G4
  0NG UK}

\author{K. Gebauer}
\affiliation{Institute for Geophysics and Extraterrestrial Physics,
  University of Braunschweig, Institute of Technology,
  Mendelssohnstra$\ss$e 3, 38106, Braunschweig, Germany}

\author{M. Hutcheon}
\affiliation{Department of Physics SUPA (Scottish Universities Physics
  Alliance), University of Strathclyde, 107 Rottenrow East, Glasgow G4
  0NG UK}
%\affiliation{Department of Physics, University of Strathclyde,
% Glasgow, United Kingdom}

\author{H. Linnartz}
\affiliation{Raymond and Beverly Sackler Laboratory for Astrophysics,
  Leiden Observatory, Postbus 9513, 2300 RA, Leiden, Netherlands}

\author{F. J. Molster}
\affiliation{Leiden Observatory, Leiden University, Postbus 9513, 2300
  RA, Leiden, Netherlands}
\affiliation{Netherlands Organization for Scientific Research,
  Council for the Physical Sciences, Postbus 93460, 2509 AL, The
  Hague, Netherlands}
%\affiliation{Leiden Observatory, Leiden University, Netherlands}

\author{B. Stoll}
\affiliation{Institute for Geophysics and Extraterrestrial Physics,
  University of Braunschweig, Institute of Technology,
  Mendelssohnstra$\ss$e 3, 38106, Braunschweig, Germany}
%\affiliation{Institute for Geophysics and Extraterrestrial Physics,
%  Technical University at Braunschweig, Germany}

\author{P. C. van der Tuijn}
\affiliation{Fijn Mechanische Dienst, Leiden Institute of Physics,
  Leiden University, Postbus 9513, 2300 RA, Leiden, Netherlands}

\author{H. J. Fraser}
\affiliation{Department of Physics SUPA (Scottish Universities Physics
  Alliance), University of Strathclyde, 107 Rottenrow East, Glasgow G4
  0NG UK}
%\affiliation{Department of Physics, University of Strathclyde, Glasgow, United Kingdom}

\author{J. Blum}%
\email[Author to whom correspondance should be addressed; electronic mail: ]{j.blum@tu-bs.de}
\affiliation{Institute for Geophysics and Extraterrestrial Physics,
  University of Braunschweig, Institute of Technology,
  Mendelssohnstra$\ss$e 3, 38106, Braunschweig, Germany}
%\affiliation{Institute for Geophysics and Extraterrestrial Physics,
%  Technical University at Braunschweig, Germany} 

\date{\today}% It is always \today, today,

\begin{abstract}
We discuss the design, operation, and performance of a vacuum setup
constructed for use in zero (or reduced) gravity conditions to
initiate collisions of fragile millimeter-sized particles at low
velocity and temperature. Such particles are typically found in
many astronomical settings and in regions of planet formation. The
instrument has participated in four parabolic flight campaigns to
date, operating for a total of 2.4 hours in reduced gravity conditions
and successfully recording over 300 separate collisions of loosely
packed dust aggregates and ice samples. The imparted particle
velocities achieved range from 0.03\,-\,0.28\,m\,s$^{-1}$ and a
high-speed, high-resolution camera captures the events at 107 frames
per second from two viewing angles separated by either 48.8$^{\circ}$
or 60.0$^{\circ}$. The particles can be stored inside the experiment
vacuum chamber at temperatures of 80\,-\,300\,K for several
uninterrupted hours using a built-in thermal accumulation system. The
copper structure allows cooling down to cryogenic temperatures before
commencement of the experiments. Throughout the parabolic flight
campaigns, add-ons and modifications have been made, illustrating the
instrument flexibility in the study of small particle collisions.
\textit{[Copyright Notice: The following article has been accepted by
    the} Journal of the Review of Scientific
  Instruments\textit{. After it is published, it can be found at
    http://rsi.aip.org ]}
 
\end{abstract}

\pacs{Valid PACS appear here}% PACS, the Physics and Astronomy


\maketitle

\section{\label{intro}Introduction}

The origin of the Solar System and the formation of planets, including
the enticing implications for the distribution of life in the
universe, are key avenues pursued in modern astronomy by observational
astronomers, theoretical physicists, and to a substantially smaller
extent, experimentalists. Given the unresolvable scale size at
interstellar distances, the embedded environment, and the uncertainty
of the dominating physical processes, it is a challenging task to
understand planetary origins when applying observation techniques and
theoretical modeling alone. Now recent progress in
laboratory experiments has yielded insight into particle collisions
during the \textit{initial} planet formation stages (e.g.~growth to
mm-sizes), complementing observational and theoretical work while
providing a repeatable and consistent recipe to describe individual
particle interactions \cite{blum2008}. In this paper, we present a new
laboratory instrument to test the \textit{next} stage (e.g.~growth to
cm-sizes) of coagulation theory, which is the leading scenario for
planet formation that describes a collisional growth mechanism. 

The scientific goals specifically target the properties of fragile,
mm-sized ice and dust aggregates during collisions. The aggregates are
analogous to the prevalent material found in proto-planetary disks
around young stars and therefore are believed to be the progenitors
for rocky Earth-like planets, as well as the core beginnings of gas
giants like Jupiter \cite{weidenschilling1993,natta2007}. But until now the
physical processes governing this crucial \textit{intermediate} stage
of collisional growth (from mm- to km-sizes) have remained largely
uninvestigated. The most challenging aspects to these
studies involve re-creating the reduced-gravity environment for the
timescales necessary to observe particle growth through collisional
sticking mechanisms. In this respect, ground-based drop tower
experiments \cite{blum2002,langkowski2008}, parabolic flight
maneuvres \cite{colwell2008}, space shuttle payload missions
\cite{colwell1999,colwell2003}, and the space-based laboratory on
board the International Space Station (ISS) \cite{love2004} offer
unique ways to counter the effects of the Earth's gravitational field
by creating a temporary condition of weightlessness.

The indirect detection of more than 300 extra-solar planets suggests
that planet formation might well be a commonly occuring process, even
a fundamental by-product of star formation. In previous laboratory and
microgravity experiments, the inital growth period for silicate dust
particles (up to mm-sizes) has already been shown to occur
quickly and effectively, due to van der Waals forces, achieving a
sticking probability of unity during particle collisions of low
relative velocities ($v_{c}$\,$\le$\,1\,m\,s$^{-1}$) that result in
larger, fluffy and porous, aggregate structures
\cite{blum2008}. Observationally, spectroscopic studies of silicate
dust emission features at 10 and 20\,$\mu$m emanating from a large
sample of circumstellar disks around young forming stars that exhibit
convincing evidence of the initial grain growth process from small,
sub-micron interstellar particles up to aggregates of several microns
\cite{kessler2006,furlan2006}. In addition, probing these disks at
longer, millimeter wavelengths reveals a slower drop-off in emission
than expected for some sources, which can be explained by a population
of particles that have already grown to millimeter sizes
\cite{lommen2007,natta2007}. 

The conclusions of both observation and experiment are consistent with
coagulation theory in these earliest stages, but the dominant physical
processes begin to change in unidentified ways as particles grow
larger and begin to settle within the circumstellar disk midplane. We
know that once the objects reach km-sizes, they possess sufficient
gravitational attraction to capture and retain additional mass,
kicking off a gravitationally-dominated period of runaway growth
leading to planetary status \cite{wetherill1989,wetherill1993}. But
how do mm-sized aggregates attain km-sizes, and thereafter runaway
growth? What are the dominant processes or conditions that allow, or
prevent, particles from growing beyond cm-sizes for which
observational diagnostics are lacking? It is this
\textit{intermediate} growth stage that may be the crux to questions
regarding the ease, efficiency, and proliferation of planets and
planetary systems in the universe. It is also here where our physical
interpretation is insufficient, and thus where we shall focus our
current experimental efforts and draw comparisons to recent
theoretical models.

At the intermediate growth stage, the particle size and the density of
particles required to guarantee a chance collision within a 4-10
second drop tower experiment, is prohibitive to a video analysis
capable of capturing an unobstructed view. Instead, a more direct
design was developed to initiate and observe individual
collisions. For this, parabolic flight was selected to preserve the
flexibility and diversity of the experiments in 22-second
intervals. Parabolic flights take place on board a specially designed
Airbus A300 aircraft and are organized regularly by the
European Space Agency (ESA) and the German Space Agency (DLR), and
they are operated by the French company, Novespace. During each flight
campaign, approximately 10-15 experiments are accomodated on 3
separate flights of 30 parabolic trajectories each. Between
parabolas, regular (1$g$) horizontal flight is maintained for several
minutes to give experimenters working alongside their setups a chance
to quickly review the data collected and to make necessary adjustments
or corrections to their setup. In total, one flight campaign offers a
combined weightless (or microgravity) period of approximately 30
minutes.

To account for the low gas densities of circumstellar disks, our
instrument uses laboratory pumps and a vacuum chamber. Previous
experimental studies give us an idea of the aggregate structure of the
mm-sized particles expected in the solar nebula from collisional
sticking \cite{blum2004b,blum2006}. Observations of young disks tell
us about the environment and composition, including the temperature
range (30-1500\,K) and the presence of ices, respectively. Our design
is able to achieve cold temperatures ($\simeq$\,250-300\,K) to probe
dust interactions, as well as cryogenic temperatures
($\simeq$\,80-250\,K) where ices may play an important role in
offering additional routes to more efficient sticking processes. In
addition, theoretical work reveals the kinematics in the layers of the
disk at these temperatures \cite{weidenschilling1993}. For the
particle sizes that we wish to study, the aggregates are just
beginning to de-couple from the pressure-supported gas that has
governed their motion up to this point and kept them suspended
higher in the disk atmosphere. This growth causes them
to sink toward the dense midplane of the disk,
giving the larger particles a non-zero relative velocity with
respect to the other (smaller) particles still coupled to the
turbulent gas. This defines the collision velocities expected, on the
order of 0.10\,m\,s$^{-1}$, for the planet-forming regions we are
studying \cite{weidenschilling1993,weidenschilling1997}. These
velocities do, however, present a great challenge to ground-based
studies where the Earth's gravitational field makes relative particle
velocities of $\ll$\,1\,m\,s$^{-1}$ very hard to achieve, and where it
may prevent, or favor, sticking processes unrealistically.

In October 2006, we proved our experimental setup during ESA's
45$^{\mathrm{th}}$ Parabolic Flight Campaign on the Novespace premises at
the Bordeaux-Merignac airport in Bordeaux, France. Additional flight
campaigns followed, including the November 2007 (ESA and DLR)
campaigns and the April 2008 DLR campaign, all also held in
Bordeaux. This paper reports on the instrument design. In Section II
we describe the separate components of the experiment, in Section III
we report on the instrument performance in microgravity, and in
Section IV we offer a short summary of some preliminary results for
the purpose of illustrating the effectiveness of the instrument with
respect to our scientific goals.

\section{Experiment Design Overview}

\begin{figure}[t!]
\centering
\includegraphics[scale=0.28]{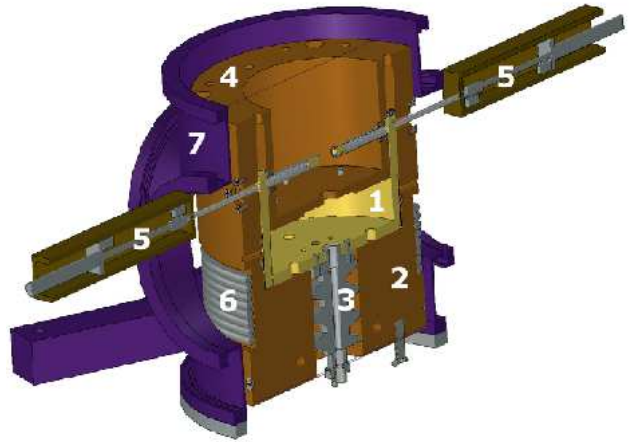}
\caption{(Color online) A CAD schematic showing a cut through the center of the
  experiment chamber. The particle storage device (1) sits on top of
  the thermal reservoir (2) built of copper. The particle storage unit
  rises up and down with a cork-screw system (3) to rotate additional
  particles into the fixed line of fire. A copper shield (4) protects
  the storage device from thermal irradiation. The two ``firing''
  pistons of the particle acceleration system stretch from opposite
  sides beginning outside the side flanges and continuing toward the
  center of the chamber (5). The whole system is chilled through
  contact with a copper tubing (6) that contains a flow of liquid
  N$_{2}$. The entire instrument is situated within a vacuum chamber
  (7). Collisions are monitored from above with the image acquisition
  system via a transparent viewport on the top flange.}
\label{cad} 
\end{figure}

The science goals driving our experiment design are the collisional
properties of fragile, mm-sized, proto-planetary dust and ice
analogs in young circumstellar disks. To obtain a large statistical
sample of this collision behavior in a simulated early proto-planetary
disk environment, a particle storage and cooling system built from
copper (Sect.~\ref{copper}) was developed to fit inside a cylindrical
vacuum chamber (of diameter 250\,mm and height 290\,mm), and to operate
over a broad range of cold (250-300\,K, for proto-planetary
\textit{dust} analogs) and cryogenic temperatures (80-250\,K, for
proto-planetary \textit{ice} analogs). A hydraulically-driven,
synchronized particle acceleration system rapidly initiated
collisions during the initial flight, but was later replaced by two
synchronized electrical DC-motors operating in a master-and-slave
configuration (Sect.~\ref{pas}). For the purpose of the experiments,
the functionality of these setups are identical in every way,
differing only in the synchronization design. Both acceleration
systems specify the velocity and trajectory of two separate dust or
ice analogs approaching from opposite directions so that the
projectiles either collide simultaneously with a removable,
dual-sided, centrally-located dust/ice target (Sect.~\ref{holder}) --
or with one another -- within the field of view of a high-speed,
high-resolution imaging and data recording system
(Sect.~\ref{das}). Finally, the vacuum chamber and the support
equipment were fit into two separate aluminum-strut support racks to
be placed on board the parabolic flight aircraft, according to safety
guidelines. A blueprint of the overall experiment design is shown in
Fig.~\ref{cad} and the following sections describe the separate
components and capabilities in greater detail. \\

\subsection{Particle Storage and Cooling}
\label{copper}

\begin{figure}[b!]
 \begin{minipage}[h!]{1.0\linewidth}
   \includegraphics[scale=0.85]{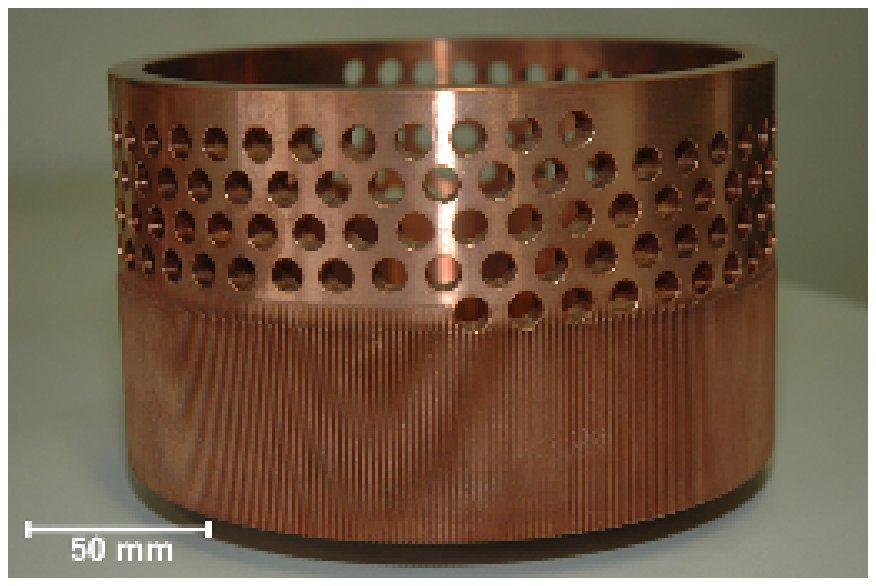}\\
   \includegraphics*[scale=1.0,viewport=0 0 105 99]{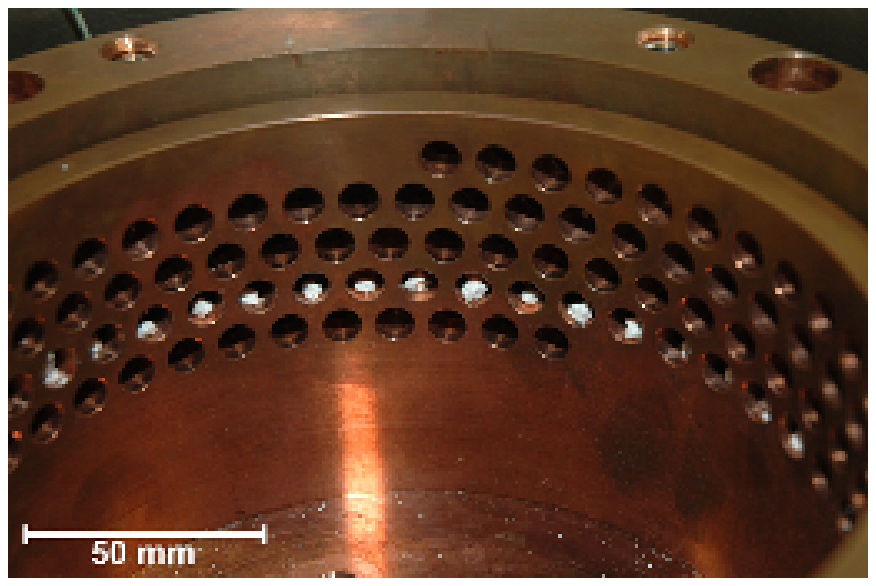}
   \includegraphics[scale=0.23]{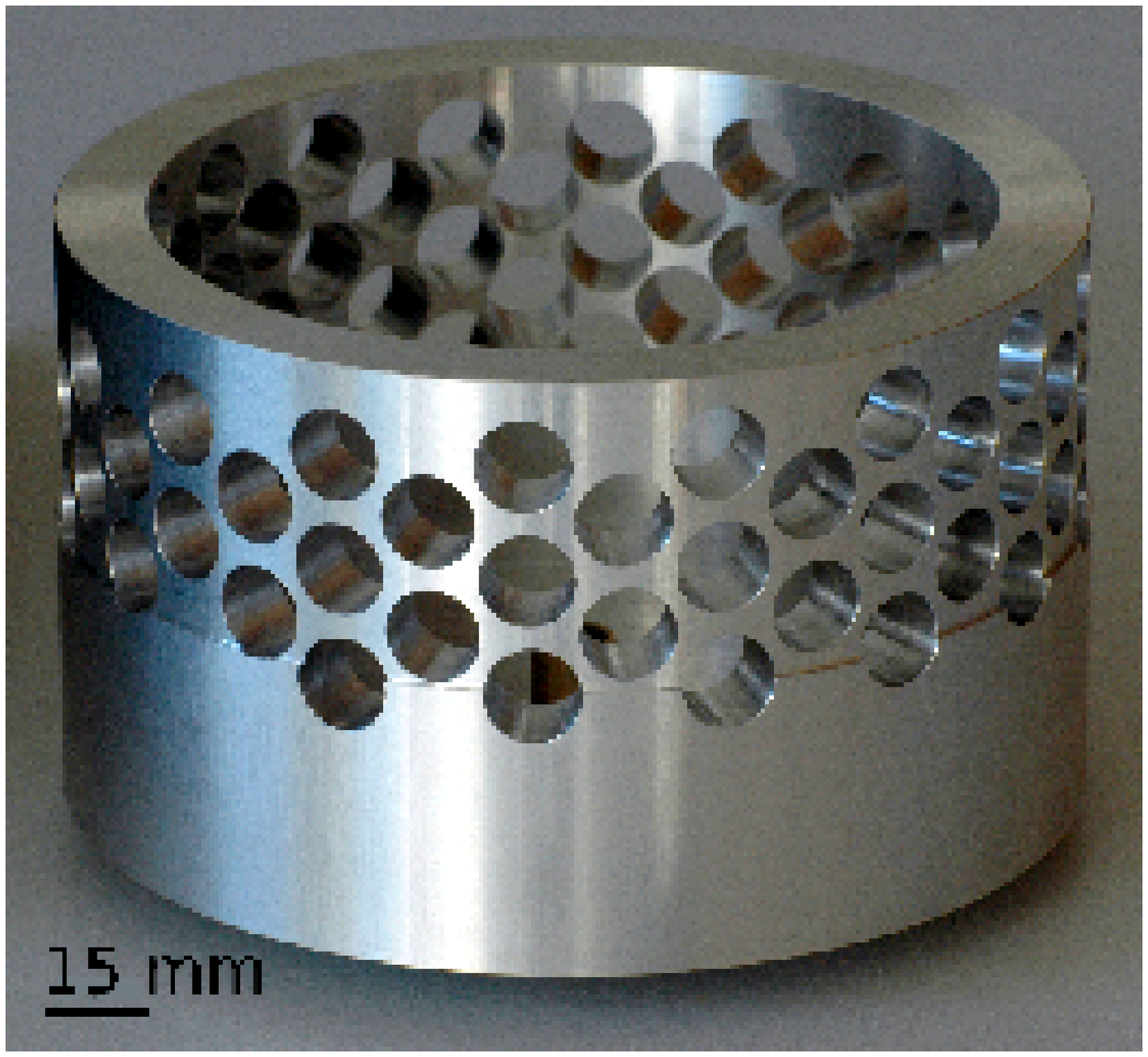}
 \end{minipage}
\caption{(Color online) At top, the original copper particle storage
  device with the storage compartments on the upper half and the
  vertical grooves to fit a gear on the lower half. Below left,
  individual compartments in one row are filled with the fragile
  mm-sized dust aggregates described in the text (see
  Sect.~\ref{performance}). At right, the aluminum version with 64
  holes for larger particle collisions. The grooves are no longer
  necessary with the revised handwheel system (see Sect.~\ref{copper}).}
\label{coliseum}
\end{figure}

An original design for particle storage was engineered
based on the requirements that the setup is capable of operating at
low and cryogenic temperatures, possesses fast and controlled
collision reloadability, obeys physical size restrictions, has maximum
sample capacity, and is in accordance with safety regulations for
parabolic flights. The result is a cylindrical storage device, or
particle reservoir, built from copper (see Fig.~\ref{coliseum}). It has a
diameter of 180\,mm and stands 110\,mm high. In the upper half of the
reservoir we drilled 180 identical, cylindrical compartments (each of
diameter 8\,mm and depth 9\,mm) situated at regular intervals
(separated by a turn of 12.5$^{\circ}$) in a double-helix pattern, such
that each hole is aligned exactly opposite a corresponding hole. When
fully loaded in this configuration, 90 separate collisions between
fragile particles (or 180 collisions of particles and central target)
can be performed. During the 2007 and 2008 follow-up flights, an
additional storage chamber of the same dimensions was conceived to
hold 64 compartments of slightly larger volume (for 32 collision
pairs). The 64 holes have a diameter of 16\,mm with a 24$^{\circ}$
turning separation. Built this time of aluminum, the new particle
storage device can be easily swapped for the original to probe the
collision properties of larger particles, up to 15\,mm.

On the lower half of the copper reservoir exterior, below the storage
compartments, vertical grooves were cut into the copper (see
Fig.~\ref{coliseum}). This way a small gear (e.g.~20\,mm in diameter
and 10\,mm high), situated to the side of the storage unit, and possessing
the same groove pattern as the storage unit, can be used to turn the
selected reservoir unit (automatically or manually) with a hand wheel
that passes out of the chamber through the bottom flange. Later, due
to difficulties operating the small gear at cryogenic temperatures
(80-250\,K), the rotation of the sample repository was revised, such
that the handwheel now directly couples to the brass-made cork-screw
system (see Fig.~\ref{cad}, label 3) with a double-Cardanic joint to
prevent canting. For this reason, a similar groove pattern was no
longer necessary on the bottom half of the aluminum version of the
storage unit.

Since cryogenic liquids and pressurized vessels are prohibited during
flight, it was necessary to devise a thermal accumulation system to
achieve, and maintain, cold and cryogenic temperatures for several
uninterrupted hours. A copper construction was deemed most suitable
based on copper's high heat capacity per unit volume, high thermal
conductivity, and ease of manufacturing for adjustments and
customization. Thus, the copper \textit{particle} reservoir is
situated on top of a 45\,kg copper \textit{thermal} reservoir, or
copper block (see label 4 in Fig.~\ref{block}). The two reservoirs
(particle and thermal) fit
together via a custom-made, quintuple thread (essentially a large
screw of diameter 30\,mm and length 70\,mm), which is firmly attached
to the bottom of the particle reservoir and fits a hollowed out groove
in the thermal block. The thread steadies the particle reservoir as it
raises out of, or lowers into, the thermal block during rotation
(supporting a height change of 5\,cm). This vertical movement allows
us to increase the storage capacity, by achieving a continuous,
winding row of samples that completes two rotations of the
reservoir. In this setup, more samples can be loaded in front of the
fixed firing system during flight, all the while maintaining thermal
contact between the storage device and the thermal reservoir for the
duration of the experiment.

Finally, a 10\,mm thin, U-shaped copper casing, or cover (label 2 in
Fig.~\ref{block}), slips over the particle reservoir on top. This
cover shields the storage device from external radiative warming
effects, and serves to ``seal'' the fragile samples in their
respective compartments, until primed for release. The cover is fixed
to the thermal block during flight, locking the (loaded) reservoir
inside and fixing the alignment. In combination with the quintuple
support thread, the slip cover helps prevent the sample repository from
tilting while still leaving enough vertical space within (above and
below) so that the reservoir can wind up and down freely. Together the
entire copper construction, shown in Fig.~\ref{block}, is then placed
within the vacuum chamber atop insulation ``feet'' constructed from
Polyetheretherketones (PEEK).

\subsection{Cryogenic Operation}
\label{cryoops}

\begin{figure}[b!]
\centering
 \includegraphics[scale=0.58]{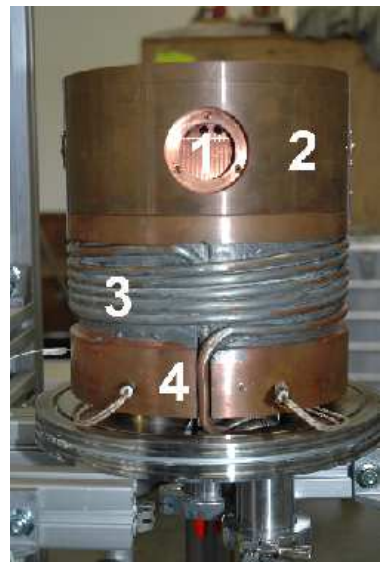}
 \caption{(Color online) The copper construction affixed atop the
   bottom flange of the vacuum chamber and isolated from the external
   environment by insulating ``feet'' constructed from PEEK. Label 1
   shows a glimpse of the reservoir via the entrance portal for the
   firing system (see Sect.~\ref{pas}); label 2 identifies the
   radiation shield; label 3 indicates the copper tubing for the LN2;
   and label 4 highlights the thermal block.} 
 \label{block}
\end{figure}

Once sealed, the vacuum chamber is pumped using the combination of a
turbo-molecular pump (TMP) and a membrane pump in series. When a
pressure of $<$\,10$^{-2}$\,mbar is achieved, cooling of the internal
environment can commence without any significant heat transfer due to
the residual gas inside the chamber. This prevents the vacuum chamber
from sweating or getting frosted on the outside. The best
vacuum attainable in this setup is 10$^{-6}$\,mbar. Liquid N$_{2}$
(LN2) enters the chamber through a LN2 feedthrough on the bottom
flange. It is connected on the inside to flexible copper tubing that
has been molded around the thermal reservoir (label 3 in
Fig.~\ref{block}). The fluid winds upwards, conductively cooling the
copper tubing, the copper block, and finally the particle reservoir
and radiation shield. K-type thermocouples monitor the system
temperature at up to 8 separate (and distributed) locations during this
process. The residual nitrogen gas and liquid that reaches the top of
the coiled tube flows back out of the chamber through the second
``exit'' port. At no point does LN2 flow freely within the chamber
itself.

Cryogenic temperatures of $\sim$80\,K can be achieved with our setup
using approximately 45\,liters of LN2, when pumped through the flexible
copper tubing for 90 minutes. Once the flow of LN2 is stopped,
the system heats up at a rate of approximately 5\,K\,h$^{-1}$ if left
unto itself under our best vacuum (see Sect.~\ref{performance}
for data taken during an experiment). After a completed series of
experiments, but before the vacuum seal is broken, the entire system
can be brought back up to 300\,K by a heat rope wound around the
copper base block, which is triggered via an external electrical
control unit. The heat rope, which is 34\,m long, has a cross-section
of 3\,mm and a heat output of 840 Watts.

\subsection{Particle Acceleration System}
\label{pas}

\begin{figure}[t!]
\centering
 \includegraphics[scale=0.08]{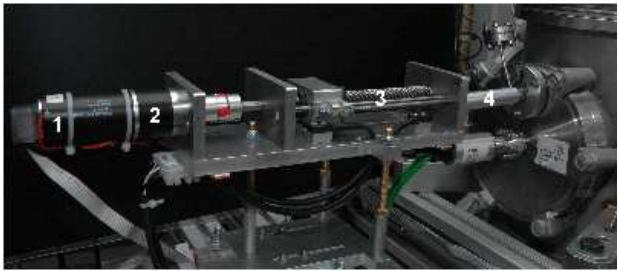}
 \caption{A close-up view of one of the two pistons in the new
   design. The labels indicate the DC-motor (1), the gear box (2), the
   stainless steel lead screw (3) with both the front and back
   switches, and finally, the vacuum feedthrough (4) with a protective
   cover. } 
 \label{pistons}
\end{figure}

To study astrophysically important collisions in this
setup, we require: very low impact velocities, constant
acceleration with no spiking due to the delicate nature of the
aggregates, clean release of the particles, and precise alignment and
synchronization of the trajectories to intersect within the camera's
field of view. Although our fluffy and porous dust aggregates are
capable of sustaining accelerations up to 100\,m\,s$^{-2}$, we limit the
maximum acceleration of the particles to 10\,m\,s$^{-2}$ in order to
prevent compaction, and thus preserve their astrophysical aggregate
design.

In the updated design, a versatile piston driving mechanism was
implemented using a master-and-slave DC-motor-system combination to
replace a hydraulically-driven setup. Two
synchronized pistons approach the center of the particle reservoir
(and chamber) from fixed positions opposite one another. The piston
back-ends extend about 0.5\,m outside the vacuum chamber and each
piston rod must pass through two separate vacuum feedthroughs before
finally breaching the internal environment. Thus, the additional
pass-through introduces an intermediate vacuum environment by applying
a chamber extension piece. This design reduces the leak rate
associated with the moving parts, as the extensions can also be pumped
in parallel, if necessary, although pumping proved nonessential in our
construction. Inside the chamber, the stainless-steel piston rods
first pass through the outer radiation shield and then through one of
the storage compartments, collecting a dust or ice aggregate on each
side with each concave-shaped piston head. The pistons accelerate the
samples towards the center of the chamber, and one another, guided
along two 80\,mm long trajectory-confining tubes, before stopping
abruptly (for particle detachment in cases of sticking) once the
samples have achieved the intended velocity.

\begin{figure}[b!]
\centering
\includegraphics[scale=0.9]{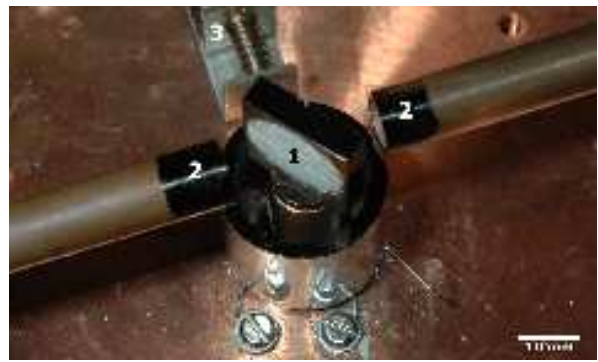}
\caption{(Color online) The dust-filled target (1) is the square piece
  at the center of the collision space. To the left and right are the
  two guiding tubes (2) within which the dust aggregates are
  accelerated. Above center, the solenoid release mechanism (3) for
  the target is visible (See Sect.~\ref{holder}).}
  \label{target}
\end{figure}

\subsection{Removable Target Holder}
\label{holder}

\begin{figure*}[th!]
 \begin{minipage}[h!]{0.45\textwidth}
  \includegraphics[scale=0.30]{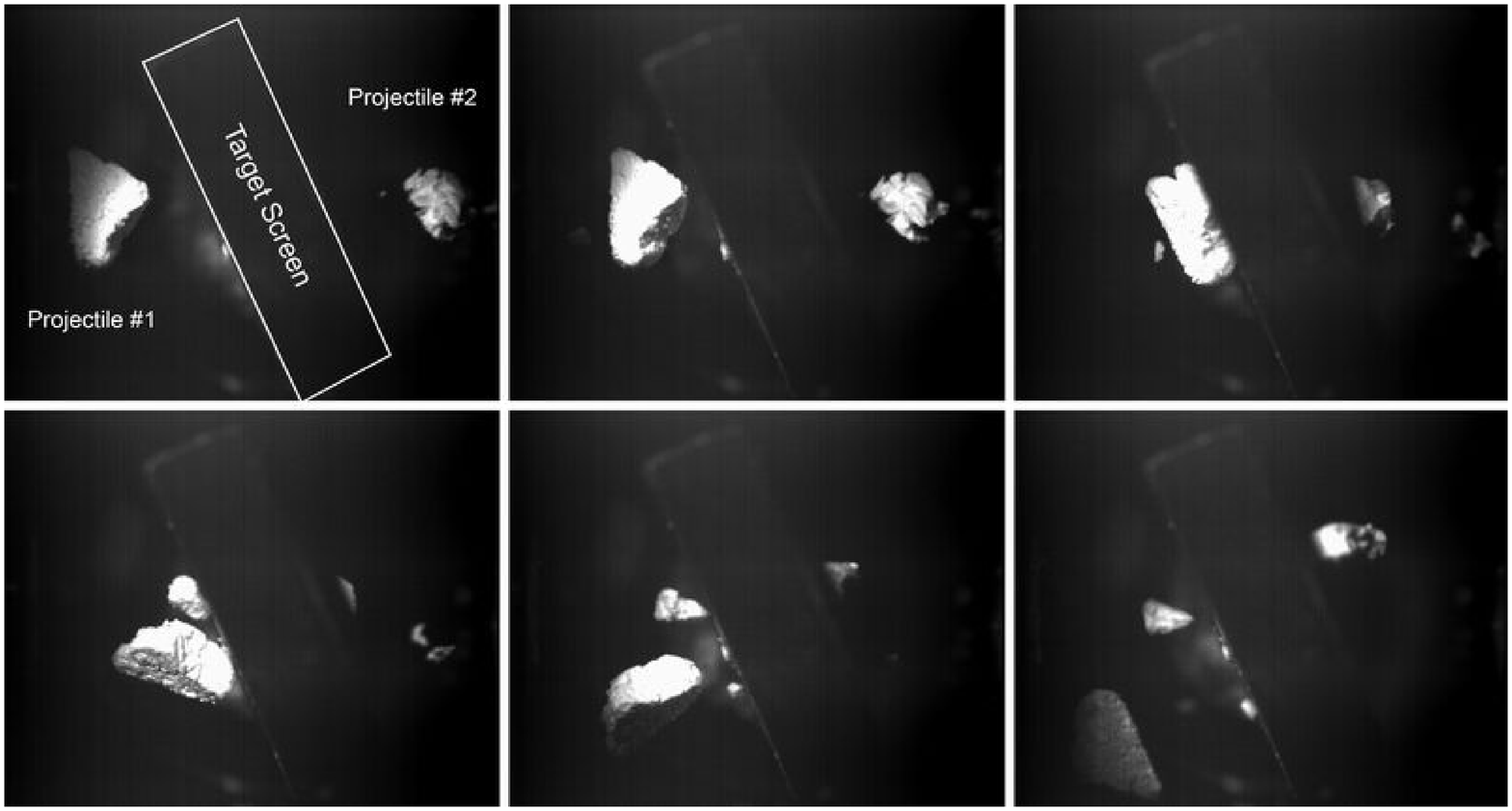}
 \end{minipage}
 \hspace{1.0cm} 
 \begin{minipage}[h!]{0.45\textwidth}
  \centering
  \includegraphics[scale=0.30]{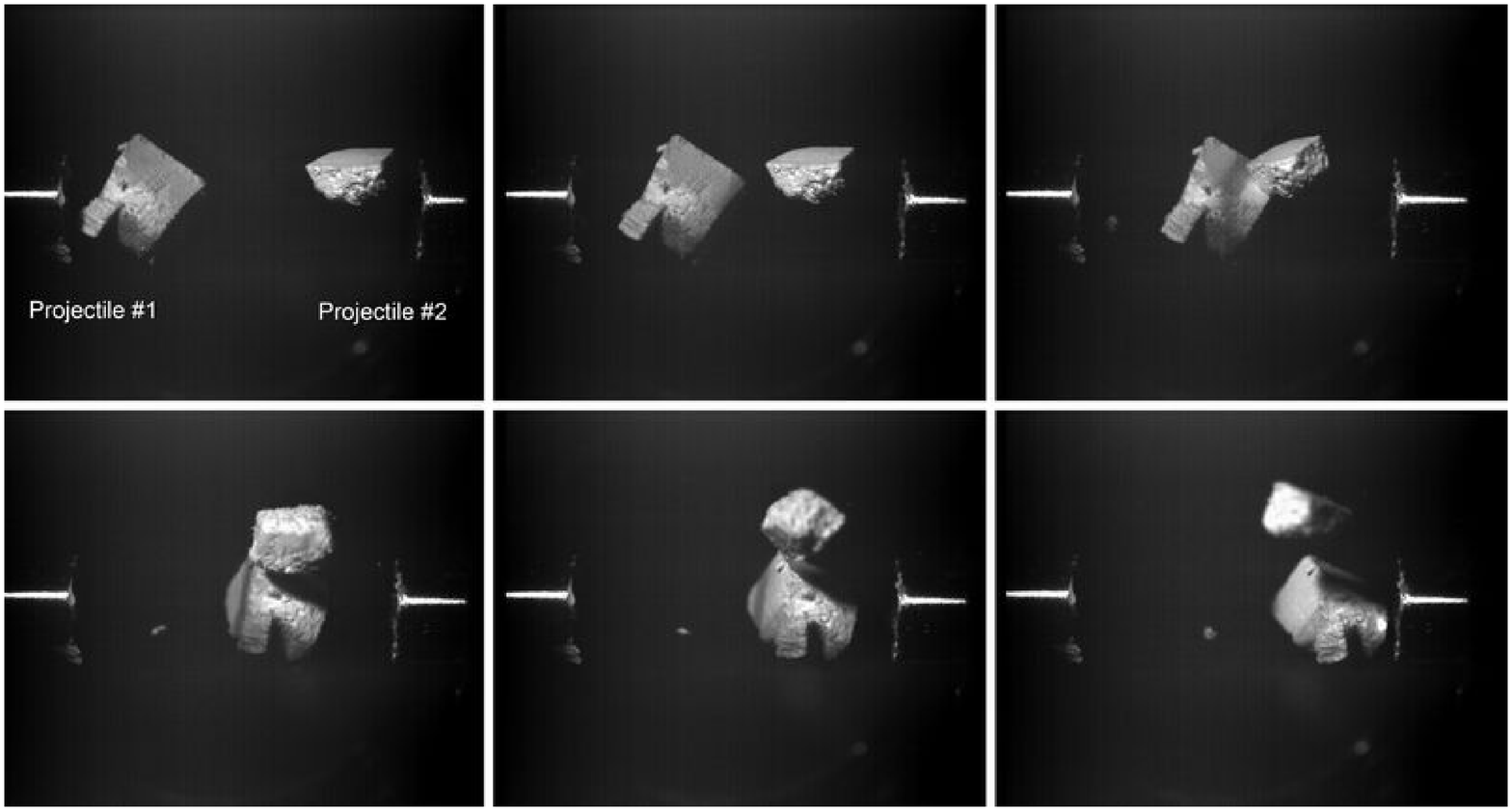}
 \end{minipage}
\caption{(Color online) Image sequences recorded during ESA's 45$^{\mathrm{th}}$
  Parabolic Flight Campaign. The sequence at left are frames
  showing aggregate-target dust collisions at a particle velocity of
  0.2\,m\,s$^{-1}$ and ambient temperature. The time-series on the
  right shows an aggregate-aggregate collision at a relative velocity
  of 0.4\,m\,s$^{-1}$. All frames measure approximately 24\,mm across
  and capture the end of the guiding tubes, which are more
  clearly visible in the right-hand sequence. In both cases, the dust
  aggregates do not stick, but instead rebound after the collision.} 
 \label{data}
\end{figure*}

To simulate impacts of our mm-sized particles against much larger
proto-planetary dust aggregates, a removable target holder was crafted
from copper to fit within the collision space between the two guiding
tubes (see Fig.~\ref{target}). The target is 17\,mm in diameter and
approximately 2\,mm deep. The dual-sided, concave structure allows us
to affix loose or compacted dust or ice samples into the mold on each
side and attach the samples with vacuum grease. The embedded target
sample is representative of a larger, possibly more rigid, compact,
and sturdier proto-planetary aggregate or particle. The target holder
itself is attached at its base to the particle storage device. This
allows it to turn throughout the experiment (synchronized with the
particle storage unit) and offers a wide range of impact angles to be
studied, from 0$^{\circ}$ (head-on) to approximately 75$^{\circ}$
(almost glancing, but before the target is edge-on and begins to block
the exits of the guiding tubes). The largest impact angles
therefore place additional limitations on the total number of
collisions that can be executed when the target is in
place. To remove this constraint at any time during the
experiment, a solenoid release mechanism allows the target holder
to drop out of the way of the collision path during normal (1$g$)
horizontal flight. Using a 24V DC signal, a lifting magnet
is triggered to pull on a small bolt that releases the two latches
that fix the target in place. Once released, and the target has
dropped out of the collision space, an additional latch fixes
the target in this lowered position until the vacuum chamber can be
safely re-opened and the target reset manually. Then, from
this point forward, only particle-particle collisions are possible.

\subsection{Data Acquisition and Imaging}
\label{das}

Capturing the collision events is achieved by a
high-speed, high-resolution state-of-the-art CMOS camera and a
recorder computer system that is capable of writing a sustained
maximum data rate of 133\,MB\,s$^{-1}$ directly to the hard disks for an
uninterrupted duration of about 33 minutes. The latter feature makes
the system indispensable for parabolic flight experiments with only
short breaks between the experiments, which do not provide sufficient
time for a read-out of imaging systems with internal memories (either
camera internal or frame grabber on-board memories). 

The camera is operated at 107 frames per second (fps) and features
1280\,(h)\,$\times$\,1024\,(v) pixels with 8 bit pixel depth (gray scale), a
pixel size of 12\,$\mu$m\,$\times$\,12\,$\mu$m, a fill factor of 40\%, and a
Base/Full CameraLink$^{\textregistered}$\,interface. The digital recording
system is a high-performance PC system that, in this special
configuration, is able to handle and store the generated data
streams of the camera. Its most prominent features comprise 2 CPUs
with 2.66\,GHz each, a PCI-X based frame grabber with 2\,GB of
internal memory, and 4 SATA-I hard disks totaling 260\,GB when bundled
as a RAID 0 array.

The camera is mounted on top of the chamber and captures the
collisions below through a transparent viewing port. Two synchronized
stroboscopic Xenon flash lamps of 1\,$\mu$s flash duration are mounted
on the top flange and provide for shadow-free illumination of the
collision volume. The flash lamps are synchronized with the camera via
a small strobe pulse adaptor box, which reverses and amplifies the
control pulses that are generated by the camera towards the
illumination source. To retrieve imaging information from multiple
viewing angles with only one camera, a 3D optics system was recently
introduced, which uses a beam splitter placed above the collision
space to capture light from two vantage points separated by either
48.8$^{\circ}$ or 60.0$^{\circ}$. The field of view covers a horizontal
grid area of approximately 24\,mm\,$\times$\,20\,mm at the focal
distance. The focal depth limits the (focused) vertical range to 5\,mm
near the collision plane. This places limitations on ground-based
testing of the setup when used for the lowest available velocity
settings.

\section{Performance in Microgravity}
\label{performance}

\begin{figure}[t!]
  \includegraphics[scale=0.35,angle=90]{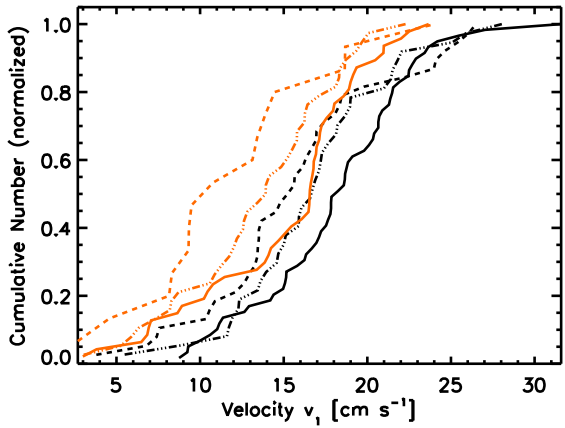}
  \caption{(Color online) The cumulative number (normalized) of all
    particles in the inaugural flight with an exit velocity
    $\le$\,$v_{1}$, using the hydraulically-driven piston setup. The
    solid, dotted, and dashed lines represent three velocity settings
    A, B, and C (intended to serve particles at speeds of 0.17, 0.19,
    and 0.20\,m\,s$^{-1}$). Black and orange lines
    differentiate between the right and left piston. Each line
    is normalized by the total number of observed particles per piston
    at that velocity setting. For the right piston, this is 59, 37,
    and 38 for setting A, B, and C, respectively. For the left piston,
    this is 47, 38, and 15 for each velocity setting. The plot
    illustrates how switching to a lower velocity setting indeed
    results in lower velocities, and that the left-hand piston
    consistently delivers particles at a slightly higher speed.}
  \label{vel}
\end{figure}
\begin{figure}[t!]
  \includegraphics[scale=0.35]{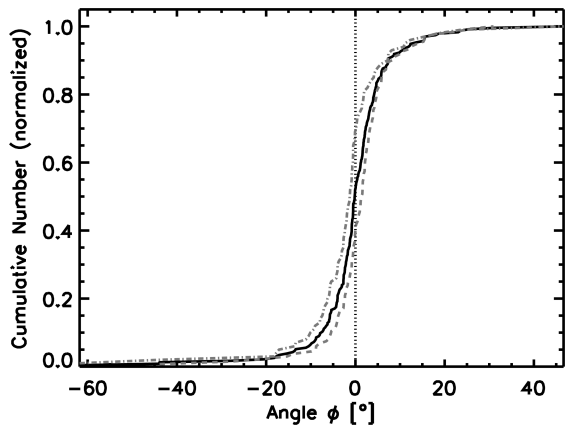}
  \caption{(Color online) The cumulative number (normalized) of all projectiles in
  the inaugural flight delivered within an exit angle
  $\le$\,$\Phi$. The solid line represents all particles whereas the
  dashed and dash-dotted curve represent the particles from the left
  and right piston, respectively. The distribution is very steep,
  showing that the angular distribution is very narrow.}
  \label{angle}
\end{figure}
\begin{figure}[b!]
  \includegraphics[scale=0.35,angle=90]{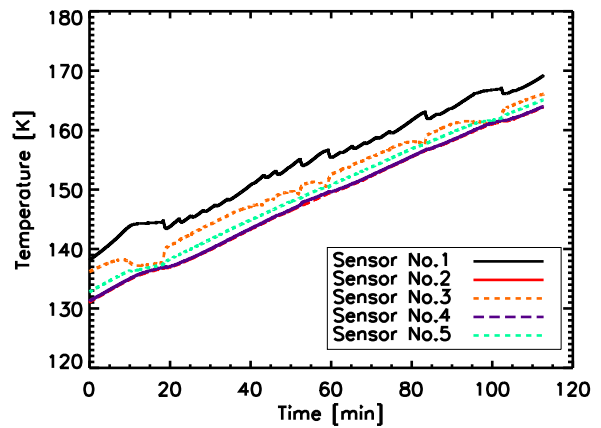}
  \caption{(Color online) The plot above shows how the temperature
    evolves during a typical parabolic flight at cryogenic
    temperatures and low (10$^{-2}$\,mbar) vacuum. Small
    fluctuations ($\pm$\,4\,K) result from different contact between
    the thermocouple and the instrument during the 0, 1, and 2$g$
    phases of the flight. Thus thermocouples located on top (\#3) or
    below (\#1) the structure will report mirrored fluctuations, while
    smaller effects are seen for thermocouples attached to the side.}
  \label{temp}
\end{figure}

The system performance in microgravity proved stable,
reliable, and accurate when flown at ambient (300\,K) and cryogenic
(130-180\,K) temperatures in vacuum. Here we will focus our
performance analysis entirely on the inaugural flight (October
2006), which was flown at ambient temperatures. Our fragile samples
were selected to simulate dust grains in the relatively warm parts
(300\,K) of the proto-planetary disk. These dust aggregates were
assembled from 1.5\,$\mu$m monodispersed monomer SiO$_{2}$ spheres
using Random Ballistic Deposition \cite{blum2004a}, measured 0.2-6\,mm
in diameter, and were 85\% porous \cite{blum2006} giving an
approximate mass range of 0.01-3.0\,g. They possessed translational
velocities of 0.10-0.28\,m\,s$^{-1}$ (the upper range of the available
experiment velocities) as they entered the collision
space, accelerated by the original hydraulically-driven piston setup. This
imparted velocity range tested during the inaugural flight was
slightly lower than expected from ground tests, but consistent
overall. Synchronization of the particle acceleration system proved
flawless and all collisions were easily observed to occur at the
center of the collision volume, and within the camera's focal plane.

We began this first experiment with aggregate and target collisions of
dust. Once we were satisfied with the resulting number of central
target collisions, we dropped the target and proceeded to study dust
aggregate-aggregate collisions. Both scenarios in this initial
experiment offer important astrophysical insight into grain growth by
probing two \textit{collisional} velocity ranges
(0.10-0.28\,m\,s$^{-1}$ for aggregate-target collisions and
0.20-0.56\,m\,s$^{-1}$, or twice the \textit{imparted} velocities, for
aggregate-aggregate collisions), as well as the influence of relative
sizes (e.g.~small projectile against a large target versus
similarly-sized colliding aggregates). Examples of the data for both
types of collision events are given in recorded image sequences in
Fig.~\ref{data}. In addition, the histograms in Fig.~\ref{vel} and
\ref{angle} show the distribution of the exit angles and velocities
for all particles with this setup, illustrating the accuracy and
consistency of particle speeds and trajectories, as designed.

We achieved 3-4 collisions per 22-second microgravity period with
manual rotation of the particle reservoir and visual verification of
the alignment. This was one collision more per parabola than initially
anticipated when designing and testing an automated system for exit
velocities of 0.10-0.28\,m\,s$^{-1}$ (e.g.~during our first and only
flight in 2006). Of course, for slower exit velocities
(0.03-0.15\,m\,s$^{-1}$) when using larger particles in the aluminum
storage unit during the follow-up campaigns, the number of possible
collisions per parabola is reduced to 1-2. Ultimately, the manual
rotation system proved far more reliable and flexible than the
autonomous design. This was largely a result of a hand-operated
turning system, which could rotate the coliseum with larger
accelerations and abrupt stops during alignment. 

Finally, the additional cryogenic functionality, as stated, also
allows for the experimentation with interstellar ice analogs, since
the abundance of ice in proto-planetary disks is high, as supported by
the high abundance of icy bodies in the outer solar system, including
the Kuiper-belt objects, icy moons, and comets. In the 2007 and 2008
follow-up flights, we initiated more than 64 additional collisions of
(larger) dust aggregates at cold temperatures (250\,K) and slower
velocities, whereby 26 clean collisions could be analyzed, and the
rest were contaminated either by third-party collisions with
previously-fired particles remaining in the collision area, or
by other complications. Finally, the experiments at cryogenic
temperatures (130-180\,K) produced more than 113 collisions of ice
particles to be analyzed.

The temperature gradients during ground-based and in-flight testing
measured just a few Kelvin across the entire copper structure, even
throughout pre-flight cooling of the experiment from 300\,K down to
80\,K via the LN2 system described in Sect.~\ref{cryoops}. However,
due to the weak pumping capabilities available during flight, the
system operated more often under a pressure of 10$^{-2}$\,mbar. This
translated to a larger warm-up rate during the cold ice
experiments, as shown in Fig.~\ref{temp} where the warming rate for a
two-hour flight was measured to be $\sim$\,15\,K\,h$^{-1}$. We note
that the change in the initial cooling temperature of 80\,K to the
130\,K that was measured at the start of the experiment
(Fig.~\ref{temp}), is due to the time between closing of the aircraft
doors and the time to reach the approved air space to perform the
flight maneuvre.

\section{Conclusions}
\label{outlook}

Our instrument design to probe the collisions of fragile particles at
low velocities certainly proved successful in its inaugural run at
ambient temperatures (300\,K) in vacuum, as well as during the
follow-up campaigns at cold (250\,K) and cryogenic (130-180\,K)
temperatures. For the ambient dust experiments in the 2006 campaign,
the 6\,mm in diameter and 85\% porous SiO$_{2}$ aggregates possessed
translational velocities from 0.10-0.28\,m\,s$^{-1}$, resulting in
particle-particle collisional energies of 1-5\,$\mu$J and
particle-target collisional energies of 0.5-2.5\,$\mu$J. The majority
of all collisions (roughly 80-90\%) resulted in semi-elastic
rebounding events, likely due to aggregate compaction. Fragmentation
occurred in about 10\% of the aggregate-aggregate collisions, but
played a much smaller role during particle-target impact events. And
finally, sticking was observed only in 10\% of the particle-target
collisions. However, the forces responsible remain unclear and warrant
additional investigation. Still, the sticking and fragmentation statistics from
this inaugural campaign suggest that we are probing a critical
transition region for the collision velocities of dust
agglomerates. And eventually it is the competition between sticking
probability, fragmentation efficiency, mass exchange, and compaction
behavior during encounters that helps shape disk lifetimes and the
effectiveness of planet formation via grain coagulation. These early
results from dust collisions at ambient temperatures exhibit the
flexibility and power of this instrument in the study of particle
collisions. Planned refinements to the experiments that have
not yet been implemented include: icy additives of amorphous (in place
of hexagonal) crystallographic structure, and the use of cryogenic
temperatures ($<$\,130\,K) and high vacuum.

\begin{acknowledgments}

We thank J.~Barrie (University of Strathclyde), M.~Costes (University
of Bordeaux I), F.~Gai (Novespace), J.~Gillan (University of
Strathclyde), E.~Jelting (Technical University at Braunschweig),
L.~Juurlink (Leiden University), M.~Krause (Technical University at
Braunschweig), A.~Orr (ESA), C.~Sikkens (Nijmegen University), the
Department of Molecular and Laser Physics of the Radboud University
for their donation of the vacuum chamber, and the members of the ESA
Topical Team: PhysicoChemistry of Ices in Space, ESTEC Contract No:
15266/01/NL/JS.

The ICES experiment was co-funded by the German Space Agency (DLR)
under grant No.~50 WM 0636, the Scottish Universities Physics Alliance
(SUPA) Astrobiology Equipment Fund, University of Strathclyde Research
Enhancement Fund, the Netherlands Research School for Astronomy
(NOVA), and the Netherlands Institute for Space Research (SRON) under
grant No.~PB-06/053. The project was generously supported by Air
Liquide, KayserThrede GmbH, Pfeiffer Vacuum, and VTS Ltd. DMS and GvdW
thank Leids Kerkhoven-Bosscha Fonds (LKBF) for assistance in attending
the first parabolic flight campaign with this instrument. We thank the
European Space Agency (ESA) and the German Space Agency (DLR) for
providing the parabolic flights.

\end{acknowledgments}

%\bibliography{ices}% Produces the bibliography via BibTeX.

\end{document}